\documentstyle[11pt,newpasp,twoside,epsf]{article}
\begin{document}
\title{The Metallicity Distribution Function of $\omega$ Centauri}
\author{Peter M. Frinchaboy, Jaehyon Rhee, James C. Ostheimer, \\ 
Steven R. Majewski, Richard J. Patterson, Winfrey Y. Johnson, \\
Dana Dinescu, Christopher Palma, Kyle B. Westfall}
\affil{University of Virginia, P. O. Box 3818, Charlottesville, VA 22903, USA}
\author{William B. Kunkel}
\affil{Las Campanas Observatory, Casilla 601, La Serena, Chile}

\begin{abstract} 

We explore the metallicity distribution function (MDF) of red giant stars
in $\omega$ Centauri from a catalogue of Washington $M$, $T_2$ and $DDO51$
photometry covering over 1.1 deg$^2$ 
outside the cluster core.  Using updated calibrations of giant branch
isometallicity loci in this filter system, photometric metallicities,
guided by previously published spectroscopic abundances,
are derived.  Several methods are employed
to correct the MDF for contamination by Galactic stars, including: 
(1) use of the surface gravity sensitivity of the ($M-DDO51$) color
index to eliminate foreground dwarf stars, (2) radial velocities, and 
(3) membership probabilities from proper motions.  The contamination-corrected MDF
for $\omega$ Cen shows a range of enrichment levels spanning nearly 2 dex in
[Fe/H], and with peaks at [Fe/H]=$-1.6$, $-1.2$, and $-0.9$.

\end{abstract}
\section{Introduction}

The large abundance spread seen in the red giant branch (RGB) of $\omega$
Cen has long been recognized as one of the unique features of this peculiar 
Milky Way globular cluster.  Recent photometric analyses
of the $\omega$ Cen RGB (e.g., Lee et al.\ 1999; Pancino et al.\ 2000, PFBPZ
hereafter; 
Majewski et al.\ 2000a, M00a hereafter) indicate a metallicity distribution 
function (MDF) stretching from 
[Fe/H]$\sim-2.0$ to perhaps as high as [Fe/H]=$-0.4$.  This spread, together with
clear evidence for a 2-4 Gyr age spread 
(Hughes \& Wallerstein 2000) as well as other unusual characteristics 
relating to its large mass, elongated shape, and internal and external dynamics
(see summary in M00a), suggests that $\omega$ Cen may represent an 
important transitional link between globular clusters and dwarf 
galaxies.  
Here we revisit the 
M00a analysis of the $\omega$ Cen MDF with the 
addition of new membership data for correcting sample contamination  
and an improved photometric metallicity calibration.

\section{Photometric Analysis}

We have imaged eight pointings of $\omega$ Cen in the Washington $M$, $T_2$ and
$DDO51$ filters with the Swope 1-m telescope and a SITe CCD 
(Fig.\ 1).  The data were reduced using standard routines, and DAOPHOT II 
and ALLFRAME 
(Stetson 1994) was used for PSF photometry.  Our resulting catalogue of stars is not
complete in the core (see Fig.\ 1) due to crowding and a conservative cut 
on stellar profile shape.  Analysis was further limited to stars with $\sigma_M$,
$\sigma_{T_2}$, and $\sigma_{DDO51} < 0.05$ mag, reducing the catalogue of
223,110 photometered stars to 52,923 stars.  Outside the core
($8\arcmin < r < 25\arcmin$), our data are complete to past the main
sequence turnoff (Fig.\ 2).

To create an $\omega$ Cen MDF, we first isolate its upper giant
branch ($M < 14.5$, $M-T_2 > 1.40$)
in the color-magnitude diagram (CMD) to limit our analysis
to where we have the greatest resolution in photometric metallicities,
and to reduce contamination from asymptotic giant branch stars
(Fig.\ 2).  This cut and the photometric error cut above combine to make 
the present analysis much more conservative than what we presented in M00a --
our goal here is to search for MDF peaks that would tend to be washed out
under the more liberal criteria used previously.
While the selected CMD region 
is dominated by $\omega$ Cen RGB stars, we want to remove
contamination by Milky Way field stars.  
The $DDO51$ filter samples the strength of the MgH+Mgb
feature at 5150\AA, which is greatly enhanced in dwarf stars (Fig.\ 3).
Via the ($M-T_2$, $M-DDO51$) diagram (Fig.\ 3b), we can eliminate most
foreground dwarf stars from the giant star sample 
(Majewski et al.\ 2000b).  Remaining
field giants are removed statistically (\S 3).

\begin{figure}
\plotfiddle{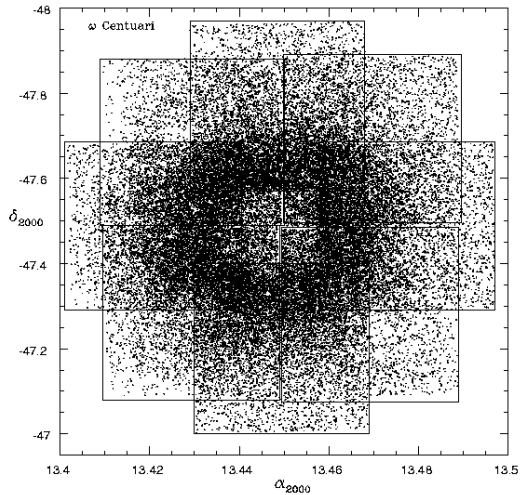}{160pt}{0}{35}{35}{-100}{-65}
\caption{Stars detected in our eight pointings of $\omega$ Cen.}
\end{figure}
\begin{figure}
\plotfiddle{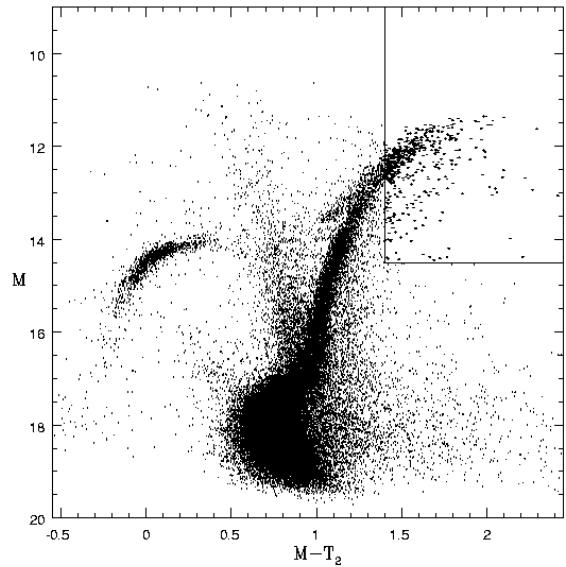}{200pt}{0}{40}{40}{-130}{-75}
\caption{The complete photometric CMD for $\omega$ Cen using the
Washington ($M$, $M-T_2$) system.  The upper right portion is the 
area of the RGB isolated for the present MDF analysis. }
\end{figure}

\begin{figure} 

\plotfiddle{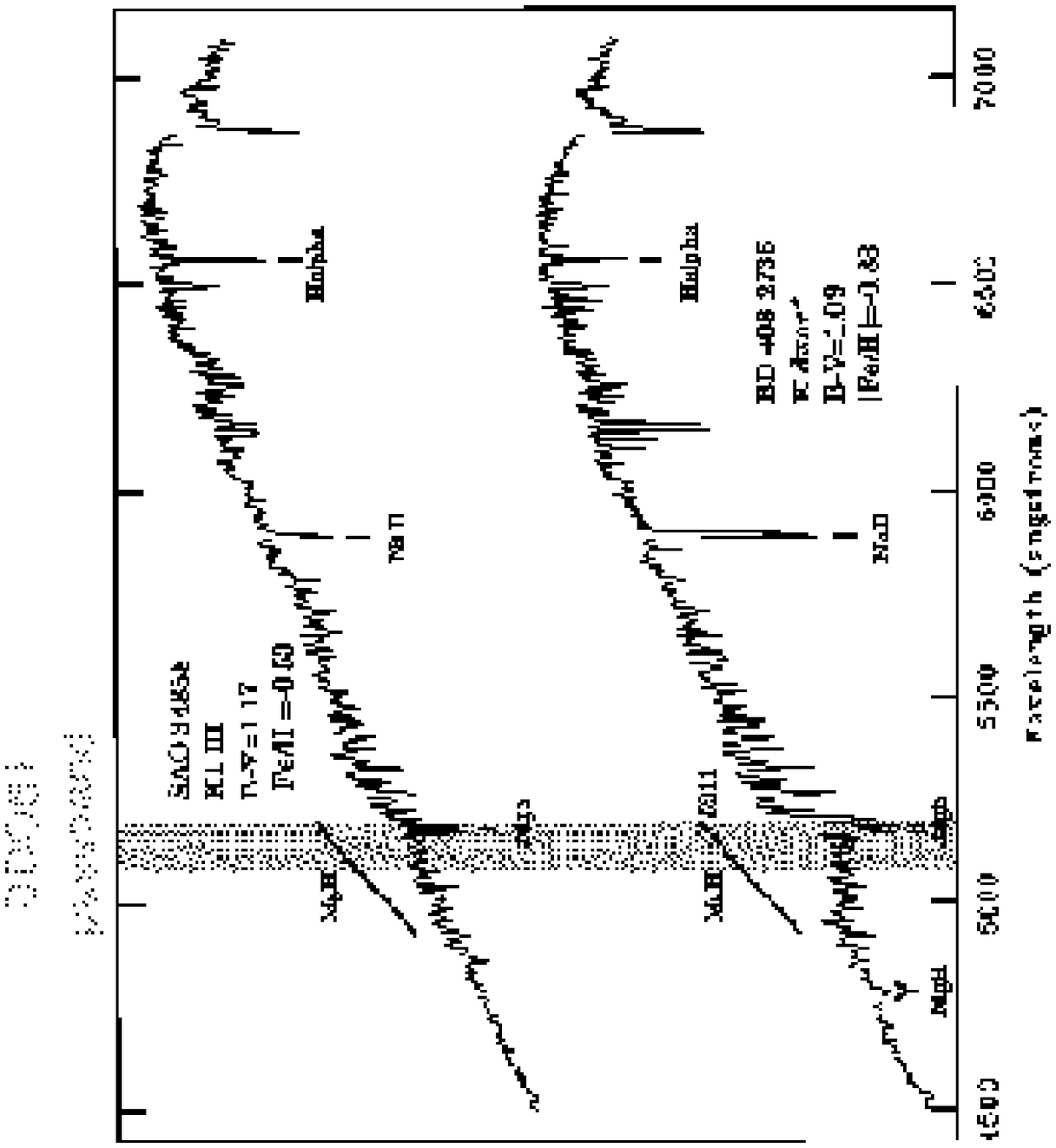}{2.0in}{270.}{33.}{33.}{-215}{160}
\plotfiddle{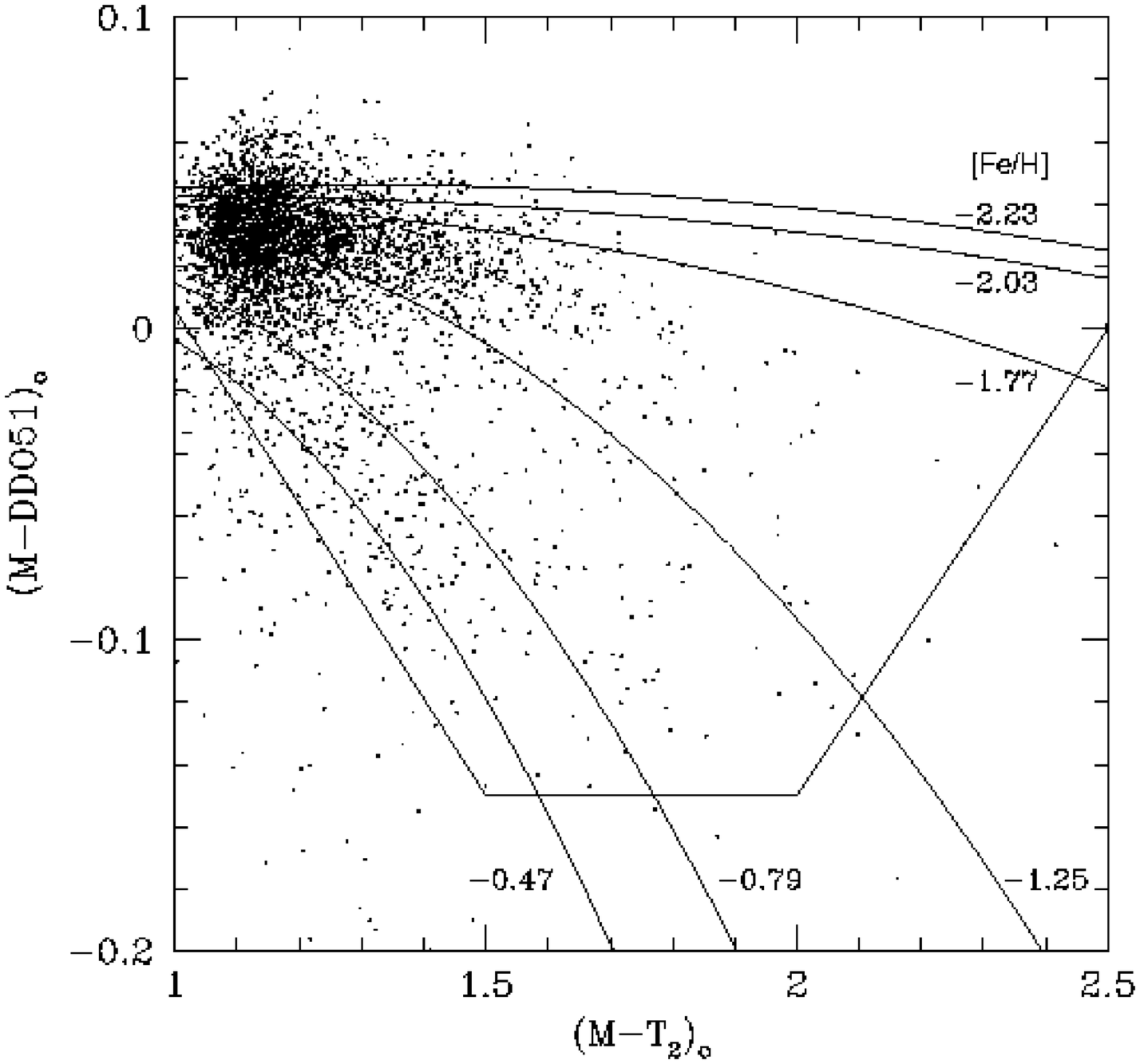}{0in}{0.}{36.}{28.4}{00}{-50}
\caption{ The $DDO51$ samples the strength of the MgH+Mgb feature which is
primarily dependent on stellar surface gravity, as may be seen in the
comparison of spectra for a K giant and K dwarf of similar metallicity
and $(B-V)$ color
(left).  Thus, $(M-DDO51)$ allows us to separate most 
dwarf stars (which
reside below the straight lines in the two-color diagram in the right
panel) from the RGB sample.  The $(M-DDO51)$ index's secondary dependence on
metallicity for RGB stars is shown by the isometallicity curves.  These
curves, when calibrated with spectroscopic data, are used to determine the
cluster MDF.}
\end{figure}

\begin{figure}
\plotfiddle{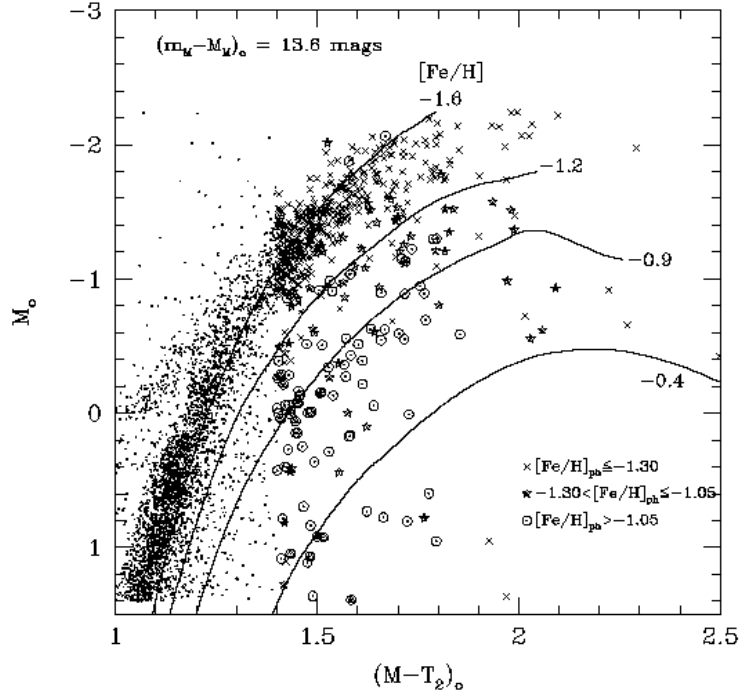}{235pt}{0}{52}{52}{-160}{-100}
\caption{$\omega$ Cen RGB stars with assigned photometric [Fe/H]
values (indicated by different symbols) from the analysis in Fig. 3. 
$Y^2$ isochrones
(Yi et al.\ 2001) at 15 Gyr (for [Fe/H]=-1.6 and -1.2) and 13 Gyr
(for [Fe/H]=-0.9 and -0.4) are provided for comparison.}
\end{figure}

\begin{figure}
\plotfiddle{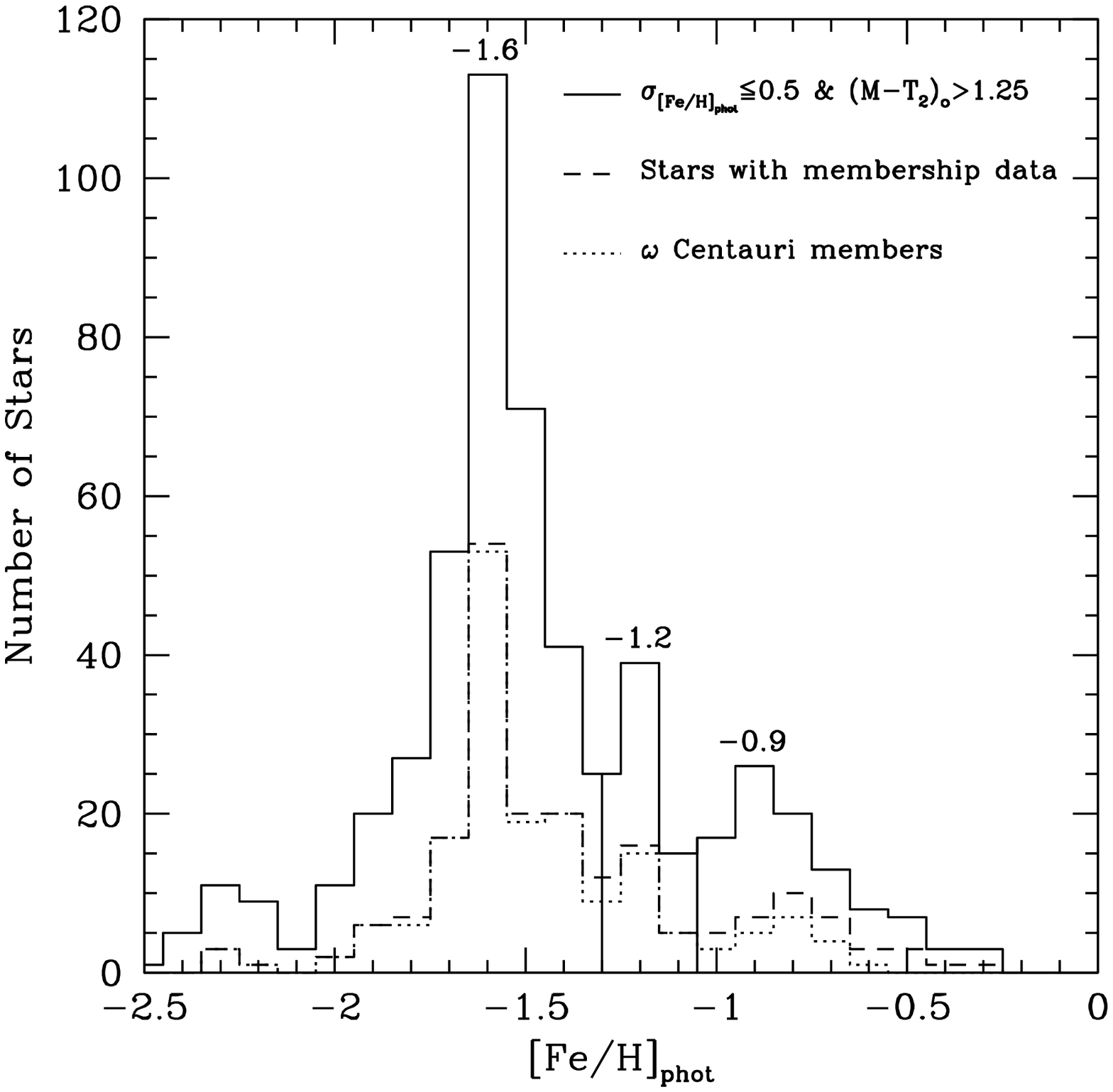}{180pt}{0}{42}{42}{-130}{-80}
\caption{The preliminary MDF derived from our photometric data (solid
line).  Also plotted are the distribution of stars having membership data
(dashed lines) and those stars that are found to be members according to
these membership data (dotted lines).  Vertical
lines demarcate the three abundance ranges shown in Fig. 4.}
\end{figure}

\section{Building the MDF}

Our MDF is based on photometric metallicities derived for
$\omega$ Cen stars as shown in Fig.\ 3.  To the extent that magnesium
tracks iron, we can derive rough [Fe/H] estimates for $\omega$ Cen
giant stars due to the secondary sensitivity of ($M-DDO51$) to
metallicity.  Isometallicity curves in the two-color diagram (Fig.\ 3) are
RGB loci from synthetic colors (Paltoglou \& Bell 1994) calibrated to fit
stars with spectroscopically determined [Fe/H] from Suntzeff \& Kraft
(1996) as discussed in Majewski et al.\ (2000b).  Stars coded by these
photometric metallicities are plotted in the RGB CMD of $\omega$ Cen in
Fig.\ 4.  $Y^2$ isochrones (Yi et al.\ 2001) for old (13-15 Gyr) giants, 
converted to the
Washington system by matching the output $T_{eff}$ and log $g$ from 
$Y^2$ to the corresponding values in the tabulated synthetic 
Washington colors by Bessell (2001), are also provided for comparison (Fig.\ 4).  
The initial MDF is constructed using these derived photometric
metallicities (Fig.\ 5).  The $Y^2$ isochrones shown in Fig.\ 4 match
metallicities that correspond to major MDF peaks in Fig.\ 5.  In general,
the derived photometric abundances for stars track the position of
isochrones of corresponding metallicity in the CMD, but of course
there is scatter.  Some scatter is due to inherent limitations 
of photometric abundances combined with observational errors, but a
significant source of scatter may derive from the intrinsic 1 dex
spread in [Mg/Fe] in $\omega$ Cen giants (Smith et al.\ 2000).
Nevertheless, we believe the data to be sufficiently reliable to
reveal gross characteristics of the MDF.

We could further limit the MDF to only known $\omega$ Cen members, 
but these are a small fraction of our data set and, moreover,
this smaller subsample is not unbiased with
respect to [Fe/H].  However, we can correct our MDF statistically
by taking advantage of the available cluster membership data to scale each
metallicity bin by the proportion of $\omega$ Cen members found among 
stars in that bin having membership data (Fig. 5).  Spectroscopic membership data 
are derived from Suntzeff \& Kraft (1996), Norris et al.\ (1996), and our 
own work.  Our spectra, centered on the calcium infrared triplet, were
obtained with the Las Campanas DuPont 2.5m $+$ ModSpec (see M00a) and the
CTIO 4-m $+$ Hydra/Loral 3k.  In total, we have obtained 68 new spectra of 
candidate RGB stars in the Fig. 4 sample, among which we identify
49 members.  Additional membership data were obtained by
matching our data to the proper motion catalogue of van Leeuwen et al.\ (2000),
and adopting as members all stars with $> 80\%$ 
probabilities from that work.  Using all membership criteria, a total of 
176 stars out of 215 stars in our Fig. 4 sample were
found to be $\omega$ Cen members (Fig. 5).  
The MDF corrected by fractional membership is shown in Fig.\ 6.

\begin{figure}
\plotfiddle{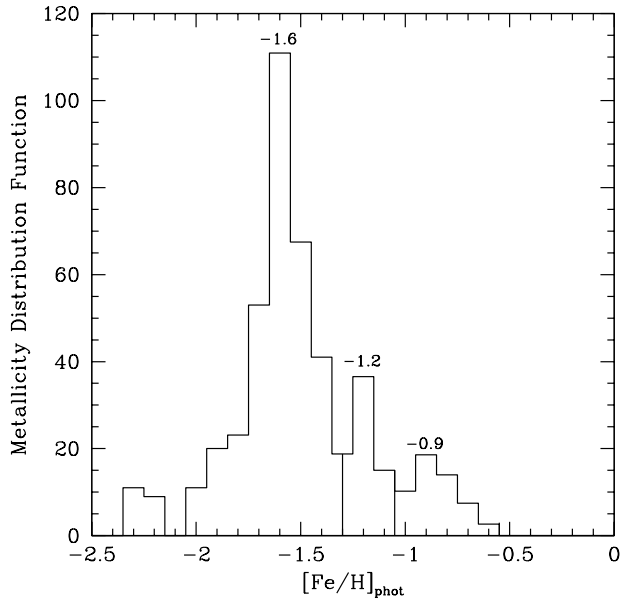}{190pt}{0}{45}{45}{-130}{-90}
\caption{The corrected MDF derived from our data normalized using membership
functions in 
Fig.\ 5.  The MDF shows
the photometric [Fe/H] peaks at $-1.6$, $-1.2$, and $-0.9$. 
Vertical lines are as in Fig. 5. }
\end{figure}

With the intention of reducing errors in derived [Fe/H],
we have adopted a much more conservative sample selection here than we used in M00a.
The result is an MDF that is less smoothly varying, with narrower, 
more defined peaks, than our previous MDF.  No doubt the width of
the peaks are still exaggerated by observational errors, but they
have now been reduced sufficiently that 
the new corrected MDF (Fig. 6) shows three distinct peaks at [Fe/H] = $-1.6$,
$-1.2$, and $-0.9$.  These peaks agree well with peaks identified by
PFBPZ at [Fe/H]=$-1.7$, $-1.3$, and $-0.8$, and by Lee et al. (1999) at [Fe/H]=$-1.7$, 
$-1.3$, and $-1.0$, 
and the overall range of [Fe/H] is consistent with the
spread found by Norris et al.\ (1996).  However, like Lee et al., while we also find evidence
for $\omega$ Cen members as metal rich as [Fe/H]$=-0.6$, Fig. 6 does not show 
an additional MDF {\it peak} at [Fe/H] = $-0.4$, corresponding to the metal
rich, ``RGB-a" giant branch identified by PFBPZ.
Surely some of the most metal rich RGB candidates from Fig. 5 may 
have been accidentally ``corrected out" of the MDF in Fig. 6 as a result
of bad luck in finding no members among the
statistically small number of stars with these metallicities available for 
the membership correction analysis.
Indeed, the impact that small number statistics 
can have on the correction process is demonstrated by the results of our
previous analysis (M00a) in which we had found almost
{\it no} $\omega$ Cen members among any star with derived 
photometric [Fe/H] $>-1.10$, which resulted in a corrected MDF terminating
near {\it that} abundance limit and completely eliminating the now obvious
$-0.9$ dex peak from that previous analysis.
But even given the lesson of this experience, the more likely reason that
we do not see as high a relative
frequency of metal rich stars in the Fig. 6 MDF compared to 
the frequency identified in the PFBPZ survey is because, as these authors
point out, their metal-rich ``RGB-a" stars are concentrated to the inner
6 arcmin of the cluster, a radius within which the present analysis faces severe incompleteness.
However, we have found that in a less conservatively selected subsample that admits 
more stars in the 
cluster core we can see the ``RGB-a" in the CMD (indeed, a trace of it can be seen 
in Figs. 2 and 4), and we have even confirmed 
radial velocity membership for eight of these metal-rich stars (almost all 
within 9 arcmin) -- unfortunately they are not admitted to the subsample under 
discussion here.  Thus, the MDF we have constructed in Fig. 6 must not 
be taken to represent the MDF of the core of $\omega$ Cen, but rather it more
closely approximates the MDF outside the core. 

We note the existence of at least one further bias in our MDF, 
which
relates to age/metallicity differences in the fractional lifetime of RGB
stars beyond our $M-T_2$ color limit.  We hope to address this in future work.

\section{Discussion: What is Omega Centauri?}

It is well known that $\omega$ Cen does not conform with most globular
clusters in a variety of ways (see M00a):  It is the most massive cluster, it shows
substantial rotation and flattening, and, of course, it has a large metallicity
spread.  Several theories about the origin of $\omega$ Cen have been
proposed, including that it is a rare cluster that (for some reason)
encountered substantial self-enrichment, that it is the product of
the merger of two stellar systems, that it is the remains of a disrupted dwarf 
spheroidal, 
and even that it derived from some amalgam of these possibilities.
That $\omega$ Cen seems to have at least {\it three} primary enrichment peaks 
and an overall [Fe/H] spread from at least $-$0.4 to $-$2.0 dex, coupled 
with claims for an age spread of up to 4 Gyr in the cluster's main sequence 
turn-off (Hughes \& Wallerstein 2000), makes a simple
two cluster merger hypothesis unlikely (see also Norris et al. 1997).
Confronted by the difficulties of multiple metallicity populations
and motivated by the relative spatial distributions of these 
populations, PFBPZ
propose a more complicated scenario -- the merger of two systems with at 
least one of the systems having undergone self-enrichment and sinking 
into the center of $\omega$ Cen.  For the merged, self-enriched entity,
which is intended to account for the two intermediate as well as the most metal rich 
populations, PFBPZ propose a giant
molecular cloud or a gas-rich protocluster.

However, a number of aspects of $\omega$ Cen lead one to suspect
its closer association with dwarf galaxies.
For example, the ``peaky" MDF of
$\omega$ Cen bears great resemblance to the burst-like, multiple
populations seen in dwarf spheroidal (dSph) galaxies (Grebel 1997).   
Interestingly, the Sagittarius (Sgr) dwarf galaxy shows a similarly large (and punctuated)
spread in [Fe/H] to $\omega$ Cen (Layden \& Sarajedini 2000).
For a variety of reasons, including the similarity 
in MDFs as well as the fact that the 
mass of $\omega$ Cen is comparable to that of the globular M54, which appears
to be the core of Sgr, it has been proposed (e.g., Lee et al.\ 1999, 
M00a) that $\omega$ Cen may be the remnant nucleus of
a tidally disrupted dwarf galaxy analogous to the Sgr system.  
As pointed out by Shetrone et al. (2001), for this model
of $\omega$ Cen formation to work, the cluster 
would have to be a daughter product of a large dwarf galaxy like Sgr, since
the heavy-element abundance patterns of 
smaller, dSph systems like Ursa Minor, Draco and Sextans 
differ from that of $\omega$ Cen, which shows 
a large enhancement of s to r-process elements with increasing 
metallicity (Smith et al. 2000).  On the other hand, the apparent greater
concentration of more metal rich stars observed in $\omega$ Cen
by PFBPZ mimics a trend seen in dwarf galaxies both great (like Fornax --
Grebel \& Stetson 1998) and small (like Sculptor -- e.g.,
Majewski et al.\ 1999). 

Apart from the actual difficulty of two clusters merging, which requires
relative velocities of $<\sim 1$ km s$^{-1}$ (Thurl \& Johnston, this proceedings), 
the merger hypothesis suffers from at least one other
unlikelihood: If $\omega$ Cen were the result of the merger of
two cluster-like systems, the parent clusters would {\it each} have to 
have been among the largest clusters in the Galaxy, and even if only 
the metal poor part of $\omega$ Cen began its life as a traditional cluster,
it too would be at the extreme end of the Galactic cluster mass scale.  
Somehow it is easier to accept that the peculiar properties of $\omega$ Cen 
are the {\it result} of its large mass, rather than that its large mass and other peculiar 
properties were accumulated as the result
of a series of unlikely occurences.  Indeed, 
the present orbit of $\omega$ Cen (i.e., barreling retrograde within and
through the Galactic plane -- Dinescu et al. 1999) 
is one that undoubtedly 
subjects it to substantial tidal stripping.  Therefore, not only was $\omega$ Cen
almost certainly larger and even more like a dwarf galaxy in the past, 
but there is every expectation  
that it has led a battered life much like its Sgr counterpart.  Evidence
for tidal tails extending from $\omega$ Cen have been reported by Leon et al.
(2000).

We have attempted to present a more accurate representation of the
MDF for $\omega$ Cen.  However, 
as pointed out by Majewski et al. (2001), 
if a system has endured substantial mass loss over its lifetime,  
one must be wary of interpreting the presently observed MDF 
to represent the true
enrichment history of that stellar system.  Older (and more extended)
populations will have had more time to have been stripped, and especially
in the case of $\omega$ Cen, whose planar orbit has almost certainly evolved
considerably, that mass loss rate may have been highly variable over the
enrichment timescale.

We thank support from the National Science Foundation, The David and Lucile
Packard Foundation, Research Corporation and Carnegie Observatories.

\end{document}